\documentclass[a4paper]{article}

\usepackage{icrc2013}
\usepackage[english]{babel}

\title{Enhancement of the radar signal of air showers due to time compression}

\shorttitle{Enhancement of the radar signal of air showers due to time compression}

\authors{
J. Stasielak$^{1,2}$,
S. Baur$^{1}$,
R. Engel$^{1}$, 
P. Neunteufel$^{1}$,
J. P\c ekala$^{2}$,
R. \v{S}m\'{\i}da$^{1}$, 
F. Werner$^{1}$, 
H. Wilczy\'nski$^{2}$.
}

\afiliations{
$^1$ Karlsruhe Institute of Technology, Karlsruhe, Germany \\
$^2$ Institute of Nuclear Physics PAN, Krakow, Poland \\
}

\email{jaroslaw.stasielak@ifj.edu.pl}

\abstract{We investigate the feasibility of detecting extensive air showers by the radar technique at viewing angles smaller than $\sim 25^\circ$ to the shower axis. Considering a bistatic radar setup and shower geometries in which the receiver points into the arriving shower, we simulate reflection of radio waves off the short-lived plasma produced by the high-energy shower particles in the air. Using the Thomson cross-section for scattering of  radio waves and summing coherently contributions of the reflected radio wave over the volume of the plasma disk, we obtain the time evolution of the signal. We analyze the spectral power density of the radar echo and the received power. Based on the obtained results, we discuss possible modes of the radar detection of extensive air showers.
}

\keywords{ultra-high energy cosmic rays, extensive air showers, radar, radio signal.}

\begin{document}
\maketitle

\section{Introduction}

The remote sensing of extensive air showers (EAS) using a bistatic radar system is a promising technique with almost 100$\%$ duty cycle that is currently being developed. If successful, it will allow the next generation of cosmic ray observatories to be built at much lower cost. 

The concept of implementing a radar for cosmic ray detection dates back to 1940 \cite{bibe:blackett}. However, due to the lack of experimental confirmation of this method, it was not pursued for several decades. In recent years renewed attention has been given to this topic \cite{bibe:baruch,bibe:gorham,bibe:bakunov,bibe:takai,bibe:stasielak} and experimental efforts to detect EAS using the radar technique were made by several groups \cite{bibe:vin,bibe:lyono,bibe:terasawa,bibe:mariachi,bibe:tara,bibe:tara2}.

Detection of EAS via radar technique is based on the principle of scattering radio waves off the plasma produced in the atmosphere by the high-energy particles of the shower. The locally produced plasma decays. For the plasma densities relevant for EAS and at low altitudes, the three-body attachment to oxygen dominates the deionization process as it depends quadratically on the oxygen density. This leads to the plasma lifetime of 10 ns at sea level and about 100 ns at an altitude of 10 km \cite{bibe:vidmar,bibe:nijdam,bibe:n}.

Some features of scattering of the radio waves from the ionization column produced by meteors or lightnings are expected to be similar to the scattering from the ionization trail left behind the shower front. Therefore, we can use these similarities as a starting-point for analysis of the radar reflection from EAS. 

The ionization trail that results from meteors or lightnings is traditionally divided  into underdense and overdense regions, depending on the local plasma frequency $\nu_p$. If the electron density is high enough that the plasma frequency exceeds the radar frequency then the radio wave is reflected from its surface. Such a region is called overdense. In contrast, if the electron density is low enough that the local plasma frequency is lower than the frequency of the incoming radio wave, then the region is underdense and the radio wave can penetrate the ionized region. In such a case the reflections are caused by the Thomson scattering of the radio wave on individual free electrons. 

Gorham \cite{bibe:gorham} considered radar reflection from the side of a horizontal ionization trail left by ultra-high energy neutrinos at an altitude of about 10 km. He suggested that the most inner (overdense) part of the ionization column is responsible for the bulk of the radar reflection. By analogy with the reflective behaviour of the overdense region produced by a meteor, he assumed that the radar cross-section (RCS) of the overdense trail produced by EAS is equal to the RCS of a thin metallic wire. 

An alternative mode of EAS detection was discussed in \cite{bibe:bakunov}, where reflection of the radar wave from the relativistically moving shower front was considered. The reflection coefficient was obtained by solving Maxwell's equations with the corresponding boundary conditions. The speed of the shower front, however, was assumed to be that of electrons moving with the speed lower than the speed of light in the air.

In reality the shower front moves with the speed of the highest energy particles in a shower and exceeds the local speed of light at all altitudes of relevance. Therefore, a reflected wave in the forward direction cannot exist because it would be immediately caught by the shower front. In its place a second transmitted wave, so called transmitted backscattered wave, is formed \cite{bibe:stephanov} and it follows in pursuit of the ionization front while standing off from it. 

It follows that, in the case of scattering of the radio waves incoming at small angles to the shower axis, one can not treat plasma as overdense because it is transparent to arbitrarily low-frequency incident radiation. Considering the plasma as underdense and using the Thomson cross-section for radar scattering seems to be justified in this situation. 
Moreover, unlike the case of reflection from the side of the ionization trail, where the frequency does not change, the frequency of the backscattered radio wave will be upshifted. One can then expect an enhancement of the backscattered signal due to its time compression. To avoid confusion we will use the term 'reflected wave' as a synonym of the scattered wave.

Scattering of the radio wave from the ionization trail produced by the EAS in the underdense plasma regime was considered in \cite{bibe:takai}. The calculations were made for the forward scattered signal assuming that the ionization occurs in a line along the shower axis, i.e. that contributions from the laterally distributed electrons are coherent. The transmitter and receiver were located 50 km apart. The computed case lies in between side scattering and front scattering with respect to the ionization column.
Thus, the frequency upshift of the received signal reaches only modest values.

In this paper, which is an extension of our previous work \cite{bibe:stasielak}, we investigate the feasibility of detecting EAS by the bistatic radar technique at viewing angles smaller than $\sim 25^\circ$ to the shower axis. Simulations are performed for the underdense regime using the Thomson cross-section for scattering of radio waves off the short-lived, non-moving plasma. We neglect absorption, multiple scattering, and currents induced in the plasma. We sum coherently contributions of the radio wave reflection on each individual electron over the volume of the disk-like ionization trail and obtain the time evolution of the radar echo. The final result depends on the individual phase factors of the scattering electrons.

\section{Modeling radar reflection}

 \begin{figure}[t]
  \centering
  \includegraphics[width=0.4\textwidth]{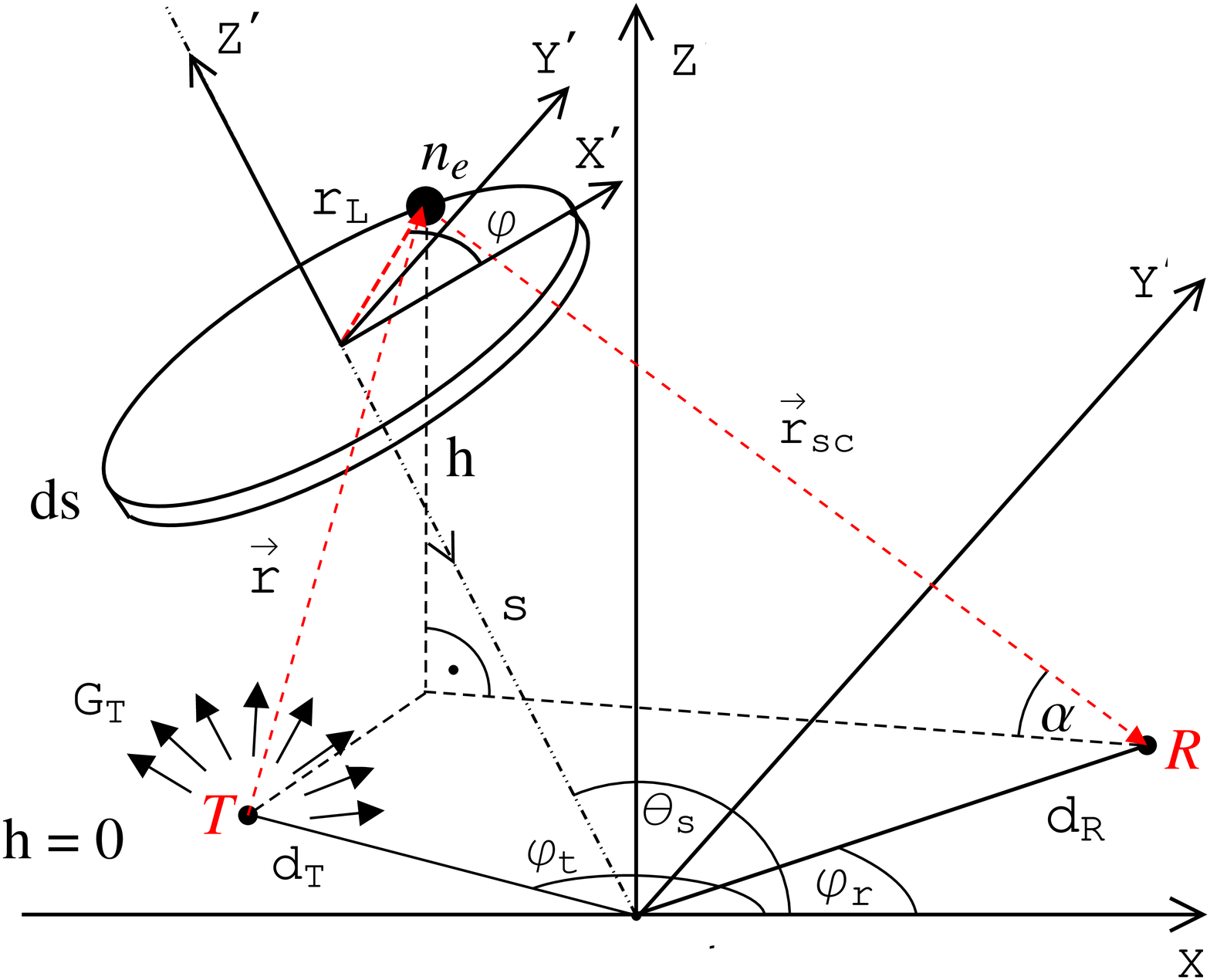}
  \caption{A schematic diagram representing a bistatic radar system and reflection from the non-moving plasma produced by the EAS in the atmosphere. See the text for a detailed explanation.}
  \label{fig:1}
 \end{figure}

 \begin{figure}[t]
  \centering
  \includegraphics[width=0.5\textwidth]{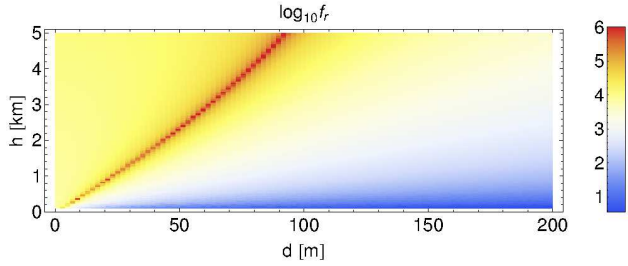}
  \caption{Distribution of the factors by which the emitted frequency is up-shifted ($f_r$) for the radio wave scattered off different parts of the shower disk. The altitude of the disk element is given by $h$, whereas its distance to the receiver in the horizontal plane is $d$. A vertical shower heading towards the transmitter is considered.}
  \label{fig:upshift}
 \end{figure}  
 
A schematic diagram representing the concept of the EAS detection using radar technique is shown in Figure \ref{fig:1}.
A ground-based radio transmitter (T) irradiates a short-lived, disk-like, non-moving plasma left behind the shower front. The radio signal is scattered by free electrons in the ionization trail and subsequently received by the ground-based antenna (R). The geometry of the bistatic radar system is determined by the polar coordinates of the transmitter and the receiver, i.e. by the distances from the shower core to the transmitter ($d_T$) and to the receiver ($d_R$) together with the angles $\varphi_t$ and $\varphi_r$. 

Let us consider the radar reflection from the disk element with coordinates ($r_L$, $\varphi$), altitude $h$, and electron density $n_e$. Its contribution to the radar echo at the receiver at time $t$ is given by 
\begin{eqnarray}
	U (t,s,r_L,\varphi)  & = & \sin \alpha \sqrt{\frac{G_T A_R}{4 \pi}} \sqrt{\frac{\mathrm{d} \sigma_T}{ \mathrm{d} \Omega}} n_e e^{i\omega t + \varphi_0}    
 \nonumber \\
           & \times &    {e^{-i\int_{{\bf r}} n{\bf k} \cdot \mathrm{d}{\bf r}} e^{ - i\int_{{\bf r_{sc}}} n{\bf k_{sc}} \cdot \mathrm{d} {\bf r_{sc}} } \over |{\bf r}| |{\bf r_{sc}}|}    
           \rm{,} \label{dUrcv}
\end{eqnarray}
where $G_T$ is the transmitter antenna gain, $A_R$ is the effective area of the receiver antenna, $\bf k$ and $\bf k_{sc}$  are the altitude-dependent wave vectors of incoming and scattered radio wave, $\varphi_0$ is the initial phase of the emitted signal, $\mathrm{d}\sigma_T/\mathrm{d}\Omega$ is the differential Thomson cross-section, $s$ is the projection of the distance between the shower core and the considered disk element on the shower axis, $\alpha$ is the inclination angle of the reflected radio wave, and $n$ is the refractive index of the air derived from the fit to the US standard atmosphere. The factor $\sin \alpha$ is included to take into account the dependence of the receiver antenna gain on the direction. We assume that the receiver is oriented vertically upwards. 

The signal received by the antenna at a given time $t$ is the sum of the signals scattered at different times and from different parts of the plasma disk. These individual contributions interfere with each other and only the integral over the whole volume $V(t)$, from which they arrive simultaneously, gives us the correct total signal. The relative amplitude of the radio wave at the receiver antenna is given by
\begin{equation}
	U (t) = \int_{V(t)} U (t,s,r_L,\varphi) r_L \mathrm{d}r_L \mathrm{d}\varphi \mathrm{d}s  \rm{.} \label{total}
\end{equation} 

The factor $U(t)$ is defined in such a way that the ratio of the 'instantaneous' power $P_R(t)$ received by the detector antenna to the power emitted by the transmitter $P_T$ is equal to 
\begin{equation}
		P_R(t)/P_T = R^2(t) \rm{,}
\end{equation}
where $ R(t)$ is the real part of $U(t)$. The term $R(t)$ is proportional to the electric field strength detected by the receiver. It is used in the Fourier analysis to obtain the power spectrum of the recorded signal. The 'real' power received by the detector $P_R$ can be obtained by averaging $R^2(t)$ (according to the time resolution of the detector), i.e. $P_R=P_T<P_R(t)/P_T>$. Note that $P_R/P_T \sim G_T A_R$. 

Alternatively, the radar reflection can be described in terms of the effective RCS, which is a measure of the target equivalent physical area of an ideal scattering surface. The effective shower RCS can be defined, by analogy with \cite{bibe:gorham}, in the following way
\begin{equation}
	\sigma (t)   =  \left| \int_{V(t)}  e^{-i\int_{{\bf r}} n{\bf k} \cdot \mathrm{d}{\bf r}} e^{ - i\int_{{\bf r_{sc}}} n{\bf k_{sc}} \cdot \mathrm{d} {\bf r_{sc}} } n_e  \sqrt{\frac{\mathrm{d} \sigma_T}{ \mathrm{d} \Omega}} \mathrm{d} V  \right|^2
 \rm{.} \label{eff-cross}
\end{equation}
Note that after removing the geometrical factor of the bistatic radar system $\frac{\sin \alpha }{|{\bf r}| |{\bf r_{sc}}|}$ from the definition of $	U (t,s,r_L,\varphi)$ we obtain $\sigma (t) \propto |U(t)|^2$.

 \begin{figure}[t]
  \vspace{-0.4cm}
  \centering
 \includegraphics[width=0.5\textwidth]{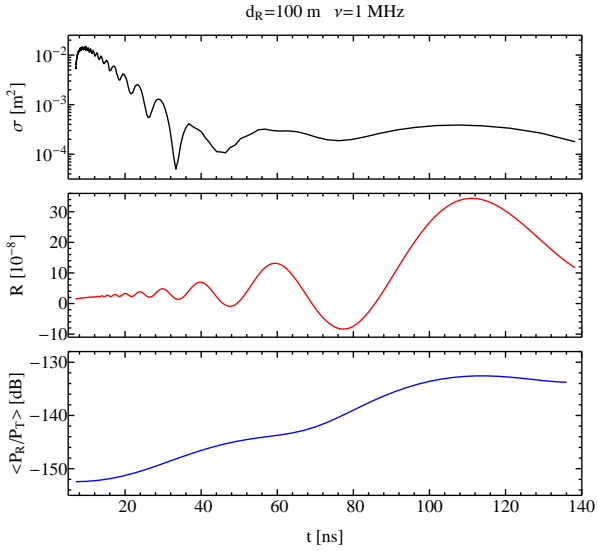}
  \caption{The effective RCS, the waveform of the radar echo, and the ratio of the power received by the detector to the emitted one (averaged over a 100 ns running time window) calculated for a vertical shower with energy $10^{18}$ eV heading towards the transmitter. The shower core-receiver distance is $d_R=100$ m and the frequency of the incident radar wave is $\nu=1$ MHz. The time $t=0$ coincides with the moment at which the shower hits the ground.}
  \label{fig:res1}
 \end{figure}

The factor $\sigma(t)$, which is defined by equation (\ref{eff-cross}), has the meaning of cross-section only when the radio transmitter and receiver are sufficiently far away from the scattering plasma or the volume of the scattering plasma is very small itself. These conditions will not always be met. In reality, the volume from which the scattered radio waves arrive simultaneously to the detector can have a considerable size for a small viewing angle to the shower axis. It is caused partly by the time compression of the reflected signal.
The estimation of the EAS radar cross-section given by $\sigma(t)$ is used only for comparison.

\section{Frequency upshift}

Despite the fact that the radio wave is scattered on a non-moving plasma, the ionization front moves with relativistic velocity, and thus we observe a Doppler shift of the received signal. Figure \ref{fig:upshift} shows the factors by which the emitted frequency is up-shifted ($f_r$) for the radio wave scattered on different parts of the disk-like plasma produced by the vertical shower heading towards the transmitter. The altitude of the plasma element is given by $h$, whereas its distance to the receiver in the horizontal plane is $d$. 

The frequency upshift depends on the wave direction and the refractive index of the air. It has the highest value for the case in which the viewing angle coincides with the Cherenkov angle. 
The typical $f_r$ is high enough to upshift a MHz signal into the GHz range. Therefore, it might be possible to observe the radar echo in the GHz range using a CROME-like setup \cite{bibe:smida} supplemented with a commercial high power MHz transmitter. 

\section{Modeling plasma and radar setup}
  
The electron density of the plasma, produced by the high-energy shower particles in the air, is estimated using the average longitudinal profile of proton showers parametrized by the Gaisser-Hillas function and assuming the Gora function \cite{bibe:gora} as the lateral distribution. We assume that each shower particle deposits on average 2.3 MeV/g/cm$^{2}$ and that all of the deposited energy goes into ionization. The mean energy per ion-pair production is 33.8 eV. We plan to improve the calculation of the plasma density by incorporating the method used in \cite{bibe:n}.

Since the received power of the radar echo is strongly diminished by the geometrical factor of $|\bf r|^{-2}|\bf r_{sc}|^{-2}$, the strongest signal will be obtained from altitudes close to the ground level, so we can assume an exponential decay of the static plasma with the characteristic time of 10 ns.

As for the bistatic radar system, we assume that the effective area of the receiver antenna is $A_R$=1 m$^2$ and the transmitter emits signal into the whole upper hemisphere (i.e. $G_T=2$).
Moreover, the receiver is ideal and its efficiency is independent of the frequency of the radar echo.

\section{Results}

 \begin{figure}[t]
  \vspace{-0.4cm}
  \centering
 \includegraphics[width=0.5\textwidth]{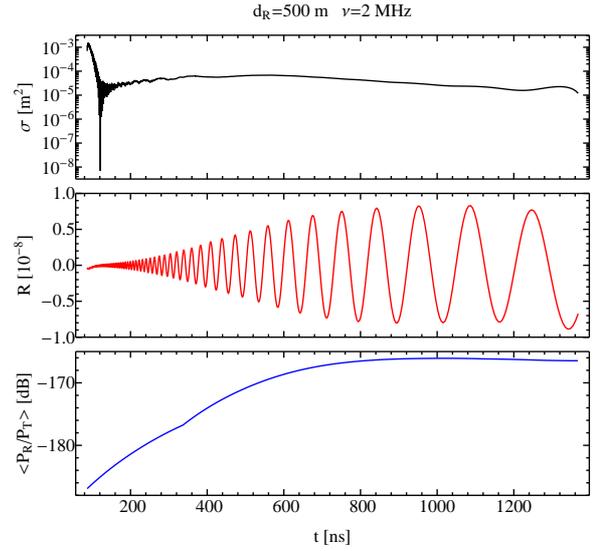}
  \caption{The same as figure \ref{fig:res1} but for the transmitter-receiver distance of $d_R=500$ m and radar frequency of $\nu=2$ MHz. The power ratio $<P_R/P_T>$ is averaged over a 500 ns running time window.}
  \label{fig:res2}
 \end{figure}

 \begin{figure}[t]
  \centering
  \includegraphics[width=0.5\textwidth]{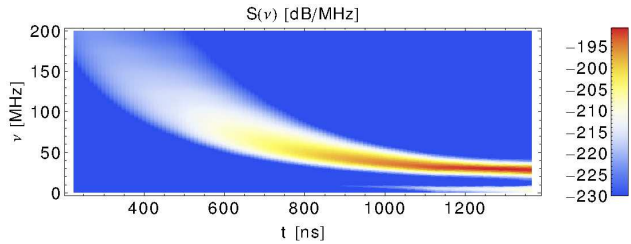}
 \vspace{-0.5cm}
  \caption{Spectrogram of the radar echo for the vertical shower with energy $10^{18}$ eV heading towards the transmitter. The radar frequency and transmitter-receiver distance are equal to $\nu=10$ MHz and $d_R=500$ m, respectively. The time window of 500 ns was used in the spectrogram calculation.}
  \label{fig:spect}
 \end{figure}
 
Figures \ref{fig:res1} and \ref{fig:res2} show the effective RCS, the waveforms $R$ of the radar echo, and the ratios of the power received by the detector to the emitted one $<P_R/P_T>$ (averaged over a running time window) calculated for different frequencies $\nu$ of the radar wave and different shower core-receiver distances $d_R$. In both cases the receiver is outside the Cherenkov cone (see figure \ref{fig:upshift}) and a vertical shower with energy $10^{18}$ eV heading towards the transmitter is considered. The time $t=0$ coincides with the moment at which the shower front hits the ground. The factor $<P_R/P_T>$, which is an equivalent of the return power, is calculated from the spectrogram of the waveform assuming an infinite bandwidth detector. Note that the waveforms are given in units of $10^{-8}$.

As we can see, the frequencies of the received signal are higher than that of the emitted one, despite the fact that the Thomson scattering preserves the frequency of the scattered radio wave. The observed upshift is caused by the interference of the radio waves reflected from the plasma volume at different stages of its development, i.e. at different times.

Since we are outside the Cherenkov cone, which is the most common case expected in experiments, the amplitudes of the waveforms increase with time. It is a geometrical effect of the reflections from the lower parts of the atmosphere. In accordance with the behavior of the signal amplitude, the return power grows with time. 
For cases when the receiver is inside the Cherenkov cone, the time sequence is reversed: the lower part of the shower is seen first and the amplitude decreases with time. 

Figures \ref{fig:res1} and \ref{fig:res2} show the enhancement of the RCS at the beginning of the signal. It is a combined effect of the increase in the size of the region from which the scattered waves arrive simultaneously to the detector and the time compression of the received signal. These two interrelated factors lead to an increase in the size of the region from which we get the coherent signal. The part of the radar echo with the highest frequency upshifts arrives to the detector from large altitudes. Therefore, despite the fact that the scattered signal is enhanced due to increase of the shower RCS, the return power of the radar echo is small due to the geometrical factor of $|\bf r|^{-2}|\bf r_{sc}|^{-2}$. 

Figure \ref{fig:spect} shows an example of the radar echo spectrogram. As expected, the frequency decreases with time. Note the low-frequency component at the end of the radar echo, which is caused by the modulation of the received signal by the factor $e^{i\omega t}$ (see equation (\ref{dUrcv})).

The values of the power ratio $<P_R/P_T>$ for different showers are given in table \ref{tab:1}. It is evident that the strength of the signal decreases with increasing radar frequency. The size of the region from which one gets a coherent signal decreases with the wavelength and destructive interference cancels out the signal from the other regions.  

\begin{table}[t]
\vspace{-0.2cm}
\caption{The maximum values of the received to the emitted power ratio for shower with different energies $E$, frequencies $\nu$ of the radar wave, and transmitter-receiver distances $d_R$. In all cases a vertical shower heading towards the transmitter is considered. The signal is averaged over a 100 ns time window.}    
\begin{center} 
\begin{tabular}{l|l|l|l|l}
\hline \hline 
\vspace{-0.3cm}
& \multicolumn{2}{|c}{} & \multicolumn{1}{|c|}{} \\
$E$ & \multicolumn{2}{c}{$10^{18}$ eV}  & \multicolumn{1}{|c}{$10^{19}$ eV} & \multicolumn{1}{|c}{$10^{20}$ eV} \\
\hline
$\nu$ \textbackslash $d_R$ & 100 m & 500 m & 200 m & 200 m \\
\hline
1 MHz & -133 dB & -158 dB & -119 dB & -98 dB \\
5 MHz & -154 dB & -174 dB & -138 dB & -117 dB \\
10 MHz & -162 dB & -182 dB & -146 dB & -124 dB \\
20 MHz & -172 dB & -191 dB & -156 dB & -134 dB \\
\hline \hline
\end{tabular}
\end{center}
\label{tab:1}   
\vspace{-0.2cm}
\end{table}

\section{Conclusions}

We have studied the feasibility of EAS detection by the bistatic radar technique at small viewing angles to the shower axis. 

Due to the time compression, the signal scattered off the plasma is enhanced. The effect is strongest for the initial part of the radar echo with the highest frequency upshifts. However, the resulting return power is strongly diminished due to the large distance of the scattering plasma to the detector.

The typical signal consists of two parts: a short signal upshifted to high-frequency with low amplitudes and a long signal with modest frequency upshifts and larger amplitudes. In principle, the last part of the signal, with the typical ratio of the received to the emitted power between -100 dB and -190 dB, should be observable. Since the strength of the signal decreases with the radar frequency, it is recommended to use low frequencies of the radar wave.

A note should be added, that the shown time traces are for an infinite bandwidth detector $-$ a realistic detector would only be able to detect the signal in a narrow frequency range. Moreover, it will only see the shower for a certain fraction of its development.

\vspace*{0.5cm}
\footnotesize{{\bf Acknowledgment:}{
This work has been supported in part by the KIT start-up grant 2066995641,
 the ASPERA project BMBF 05A11VKA and the National Centre for Research and Development (NCBiR) grant
 ERA-NET-ASPERA/01/11.}}

\end{document}